# Low-Frequency Radio Bursts and Space Weather*

N. Gopalswamy
Solar Physics Laboratory, Heliophysics Division
NASA Goddard Space Flight Center
Greenbelt, Maryland, USA
nat.gopalswamy@nasa.gov

*Abstract*—Low-frequency radio phenomena are due to the presence of nonthermal electrons in the interplanetary (IP) medium. Understanding these phenomena is important in characterizing the space environment near Earth and other destinations in the solar system. Substantial progress has been made in the past two decades, because of the continuous and uniform data sets available from space-based radio and white-light instrumentation. This paper highlights some recent results obtained on IP radio phenomena. In particular, the source of type IV radio bursts, the behavior of type III storms, shock propagation in the IP medium, and the solar-cycle variation of type II radio bursts are considered. All these phenomena are closely related to solar eruptions and active region evolution. The results presented were obtained by combining data from the Wind and SOHO missions.

## I. Introduction

All solar radio bursts are produced by the interaction of nonthermal electrons with the plasma and magnetic field in the ambient medium. Radio bursts not only provide information on the disturbances that cause them, but also on the properties of the ambient medium because the radio emission depends on the ambient density, magnetic field, and the level of turbulence. The mechanisms that accelerate electrons are also able to accelerate ions. Thus, radio bursts of various kinds are also useful indicators of solar energetic particle (SEP) events in the heliosphere that are important for space weather. Nonthermal electrons injected toward the Sun from the corona produce microwave bursts, hard X-rays, and gamma rays. Microwave bursts at frequencies similar to those of Global Positioning System (GPS) signals can occasionally drown those signals, causing communication disruptions [1-2]. The decrease in the signal-to-noise ratio in the GPS receivers located on the sunlit hemisphere of Earth is proportional to the amplitude of the solar microwave burst. Nonthermal electrons traveling away from the Sun along open field lines produce type III radio bursts and storms over a wide range of wavelengths [3]. These are generally lower energy electrons ($\leq$ 10 keV). Electrons accelerated at the flare site and trapped in moving and stationary magnetic structures produce type IV bursts. Electrons that are accelerated in shocks driven by coronal mass ejections (CMEs) produce type II radio bursts, and hence useful in obtaining information on shocks near the Sun and in the IP medium. Shocks can form as close as ~1.2 solar radii (Rs) from the Sun center [4]. Shocks also accelerate SEPs that can cause a wide variety of space weather effects in IPspacecraft, airplanes in polar routes, in the polar ionosphere, and in the atmosphere [5]. When shocks arrive at Earth, they can cause the sudden commencement of geomagnetic storms and energetic storm particle events.

I confine to low-frequency radio bursts because they represent disturbances leaving the Sun permanently, and hence are highly relevant for space weather [6-7]. The set of issues discussed here represent progress made by combining radio and white-light observations. The results presented in this paper are mainly obtained using data from the Solar and Heliospheric Observatory (SOHO) and the Wind missions. In particular, we use the images from the Large Angle and Spectrometric Coronagraph (LASCO) on board SOHO [8] and the Radio and Plasma Wave (WAVES) experiment [9] on board Wind.

## II. LOW-FREQUENCY TYPE IV BURSTS

The source of energy for the IP type IV bursts has been controversial. A flare blast wave propagating behind the associated CME, accelerating electrons, and injecting them into the CME loops has been suggested as a possibility [10]. However, the limited frequency extent of the type IV bursts and the directivity of the emission along narrow cones overlying the flare site point to the possibility of electrons accelerated at the flare site and injected into tall flare loops [6]. There is also additional evidence that the low-frequency type IV bursts occur only during the decay phase of the associated soft X-ray (SXR) flares as shown in Figs. 1 and 2. While the onset is right after the SXR peak, the end invariably coincides with the end of the SXR flare. The 2005 January 15 event was associated with an M8.6 flare from close to the disk center (N16E04). In disk events (central meridian distance < 60°) the type IV burst appears complete—starting at the highest WAVES frequency, descending to a low frequency and receding back to the highest frequency. In limb eruptions, as in the case of the 2003 November 3 event from N08W77 shown in Fig. 2, the type IV burst is partial in that only the descending part is observed. It is also worth pointing out that the bursts shown in Figs. 1 and 2 are not moving type IV bursts. Moving type IV bursts are thought to be originating from moving magnetic structures associated with CMEs, so the sources must be more extended. The distinct difference in the appearance of spectra in disk-center and limb events [11] should not be observed if the source is extended. Fig. 3 shows the source locations of a large number of type IV bursts observed during solar cycle 23. We see that the number of events drops rapidly towards the limb, and the few limb events had distinct spectra. These observations confirm that the bursts are not extended, but rather confined to a narrow cone.

*Work supported by NASA's Living with a Star TR&T program.

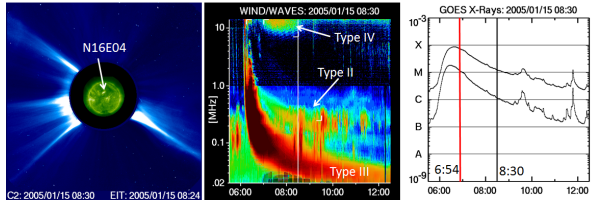

Fig.1. (left) Source location of the 2005 January 15 type IV burst (N16E04) from SOHO. (middle) Radio dynamic spectra from Wind/WAVES showing the type IV burst, accompanied by type III and type II bursts. (right) GOES soft X-ray light curves in the 1-8 Å (upper curve) and 0.5-4 Å (lower curve) channels. The vertical red and black lines mark the onset (6:54 UT) and end (8:30 UT) of the type IV burst at ~14 MHz.

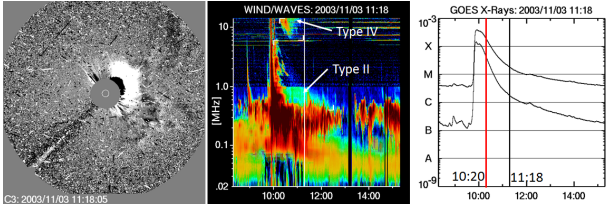

Fig.2. (left) Snapshot of the white-light CME from SOHO/LASCO originating from N08W77. (middle) Radio dynamic spectra from Wind/WAVES showing the type IV burst. Note that the type IV burst looks partial, which is typical of limb eruptions. (right) GOES soft X-ray light curves with the onset (10:20 UT) and end (11:18 UT) of the type IV burst at ~14 MHz marked.

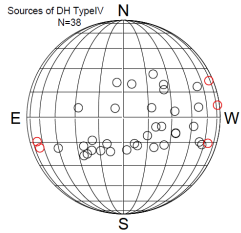

Fig.3. Source locations of 38 type IV bursts at decameter-hectometric (DH) wavelengths observed in cycle 23. Black and red circles represent disk and limb events. There were only 5 limb events.

### III. TYPE III STORMS

Noise storms represent non-eruptive energy release in active regions that result in large clusters of very short duration type I and type III bursts [12]. Type I storms typically happen at metric wavelengths, while type III storms happen at decametric and longer wavelengths [3]. Type I storms transition to type III storms at longer wavelengths suggesting that electrons gain access to open field lines [13]. One of the interesting properties of type III storms is that the number of bursts increase as the source active region crosses the central meridian. When a large CME erupts in an active region with an on-going type III storm, the storm is disrupted but returns in several hours [14]. We consider one such active region.

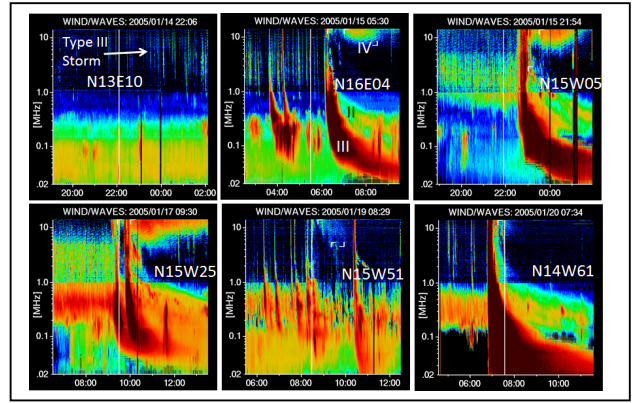

Fig. 4. A type III storm associated with active region (AR) NOAA 10720. The storm started on 2005 January 14 when the AR was at N12E10 and ended on January 20. The five major eruptions that disrupted the storm were accompanied by energetic CMEs, type II, type III, and type IV bursts. The cessation of the storm can be readily seen by comparing the radio noise to the left and right of the intense type III bursts.

Figure 4 shows a type III storm starting on 2005 January 14. The associated active region emerged on the front side of the Sun on 2005 January 11 as a bipolar region (NOAA AR 10720, N09E51). The active region grew and attained some complexity on January 14 (N13E10), when the type III storm started. The type III storm was repeatedly disrupted by CMEs, the last big one being the January 20 CME, an extreme event. The Wind/WAVES dynamic spectrum in Fig. 4 shows the disruptions. After the January 20 CME, the storm did not come back. On January 21, the AR was already at W84.

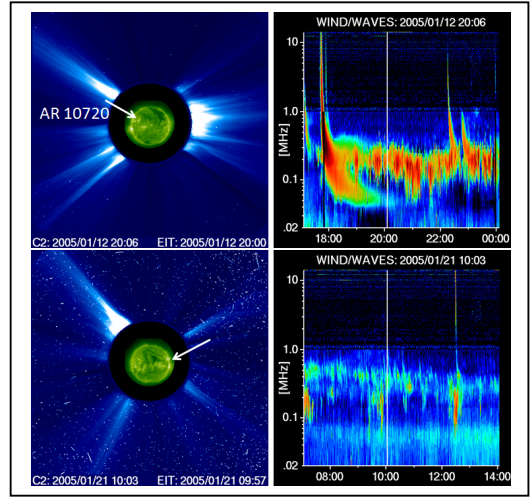

Fig. 5. Coronagraph image and the radio dynamic spectrum: (top) on 2005 January 12 before the start of the type III storm, and (bottom) on 2005 January 21 after the end of the storm. In the radio spectra, there is no storm; only auroral kilometric radiation is observed at frequencies below 1 MHz. The arrows point to the active region on SOHO's EUV images superposed on the coronagraph images.

Fig. 5 shows that the dynamic spectra on January 12 and January 21 were free of type III storms. It is not clear if the disappearance of the storm is due to the complete reconfiguration of the active region, the directivity of the storm, or the AR complexity. It appears that all these may be factors. This question can be addressed using STEREO observations because an active region rotating behind the west limb can be tracked by the STA spacecraft. A practical application of the storm disruption by a solar eruption is that the associated CME must be front-sided and likely to be close to the disk center and hence potentially a geoeffective event.

## IV. A NEW SIGNATURE OF CME INTERACTION

Interaction between CMEs have been recognized in the coronagraph field of view with a simultaneous spectral feature in the Wind/WAVES dynamic spectrum as an enhancement in the type II radio burst [15]. The coronal region probed by the two instruments have very good overlap. CME interactions have important implications for SEP events [16], CME travel time [17], and the CME/shock arrival at Earth [18]. Here we report on a new spectral feature in the radio dynamic spectrum indicating CME interaction on 2013 May 22. This event was well observed [19] and the implication for SEPs has already been reported [20].

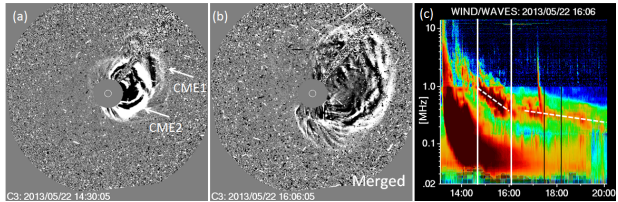

Fig. 6. (a) SOHO/LASCO difference image at 14:30 UT showing CME1 and CME2 before merger. (b) the resultant CME at 16:06 UT after merger. (c) Wind/WAVES dynamic spectrum showing the type II burst. The two vertical white lines mark the interval of type II intensification. The slope of the type II changed at the end of the intensification as shown by the dotted lines.

Fig. 6 shows two snapshots of the interacting CMEs and the associated type II radio burst. CME1 was the preceding CME with a speed of ~700 km/s followed by CME2 with a higher speed (~1450 km/s). When CME2 approached CME1, the type II burst intensified as is well known [15]. The intensification continued for ~1.5 h, until the two CMEs completely merged. After the merger, the intensification stopped, but the slope of the type II burst suddenly changed from a high value to a low value at 16:06 UT. It appears that the resultant CME slowed down after the merger to a speed of ~1310 km/s. We compare CME kinematics before and after the merger to understand the slope change.

The average frequency (f) drift rate df/dt dropped by an order of magnitude from $1.4 \times 10^{-4}$ MHz/s before merger to $1.4 \times 10^{-5}$ MHz/s after. For an isolated type II burst emitting at the fundamental plasma frequency, one can get the shock speed $V_s$ from df/dt and the density scale height $L = [(1/n)dn/dr]^{-1}$, where n(r) is the electron density as a function of the heliocentric distance (r): $V_s = 2L(1/f)(df/dt)$. Emission frequencies below 1 MHz correspond to the interplanetary medium where $n \sim r^{-2}$, so the density scale height $L=r/2$ and $V_s = (r/f)(df/dt)$. At the midpoint of the intensification around 15:18 UT, f = 0.6 MHz and the CME leading edge was at r = 18.7 Rs, giving $V_s$ = 3050 km/s. This speed is more than twice too large compared to the CME speed. The discrepancy can be readily attributed to the fact that the scale height L=9.35 Rs is not the true scale height because the density variation is not smooth. CME1 was located at a distance of ~4.5 Rs ahead of CME2, so the effective scale height becomes 4.5 Rs, which gives $V_s$ = 1468 km/s in good agreement with the CME speed. After the merger, the resultant CME propagated through an undisturbed medium, so the normal scale height applies. At 18 UT, f=0.3 MHz and the extrapolated CME distance r = 40 Rs. Since df/dt = $1.4 \times 10^{-5}$ MHz/s, we get $V_s$ = 1306 km/s, again in good agreement with the speed of the resultant CME.

This event illustrates the powerful radio tool to get the shock speed in the IP medium, except in cases when a large density structure lies ahead of the shock. When the CME observation is not available, one can still derive the speed using a density model (see e.g., [21]) that can be approximated by the inverse-square density variation. Some workers have started using the dynamic spectra of type II bursts to measure the shock speed far away from the Sun with significant success [22].

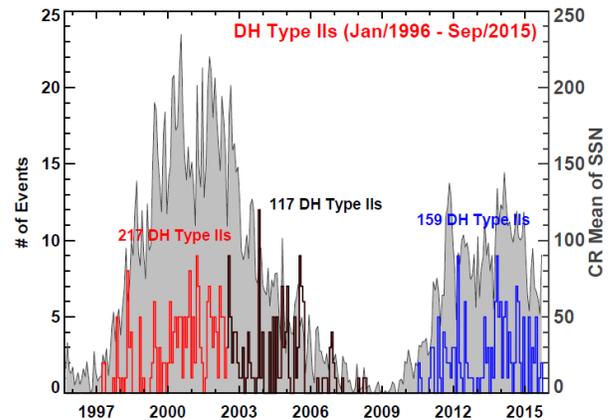

Fig. 7. Variation of the number of DH type II bursts binned over Carrington rotation (CR) periods. The red and blue curves represent the first 80 months of cycles 23 and 24, respectively. The black curve represents type II bursts during the remaining part of cycle 23. The sunspot number (averaged over CR periods) is shown for comparison in grey.

## V. SOLAR CYCLE VARIATION OF IP TYPE II BURSTS

Type II bursts in the decameter-hectometric (DH) range are good indicators of shocks in the IP medium and hence are useful in isolating shock-driving CMEs that are relevant for space weather [23]. It is important to note that only a few percent of all CMEs drive shocks and therefore isolating these CMEs using their ability to drive shocks greatly facilitates space weather prediction. For SEP events, one has to have a good magnetic connectivity to Earth, so roughly half of the DH type II bursts (from the western hemisphere) are associated with SEP events at Earth [24]. The sunspot number (SSN) has been traditionally used as an important indicator of the severity of solar activity and the resulting space weather events. Since sunspot regions can store and release large amounts of magnetic energy, it is natural to expect energetic CMEs originating from these regions and hence a good correlation between SSN and the number of DH type II bursts. Figure 7 shows the solar cycle variation of the number of DH type II bursts binned over Carrington rotation periods. There is an overall correlation between SSN and the number of DH type II bursts. However, there is a clear inter-cycle variation in the relationship: (1) The number of DH type II bursts per unit SSN is substantially higher in cycle 24 (4.7/SSN compared to 3.7/SSN in cycle 23). (2) There is a 40% drop in SSN in cycle 24 from cycle 23 over the same epoch, while the number of DH type II bursts dropped only by 27% from 217 to 159. DH type II bursts are associated with fast (speed $\geq$ 900 km/s) and wide (width $\geq$ 60º) or FW CMEs. From the online CME data base (http://cdaw.gsfc.nasa.gov) we counted 211 FW CMEs over the first 80 months of cycle 24. Over the same epoch, there were 263 FW CMEs in cycle 23. Accounting for the 5-month data gap in cycle 23, this number is more likely 280. The drop in the number of FW CMEs is therefore 25%, similar to that in the number of DH type II bursts. In other words, there is a much closer relationship between DH type II bursts and energetic CMEs because these CMEs are likely to drive shocks, which are responsible for type II bursts. It was recently shown that about 13% of cycle-24 large SEP events were associated with filament eruptions outside of active regions [25]. All these SEP events were associated with FW CMEs and DH type II bursts. This explains the discordant behavior of SSN and the number of FW CMEs (and hence DH type II bursts). The higher abundance of DH type II bursts (per SSN) in cycle 24 can be attributed to the decrease in the ambient Alfven speed in cycle 24, which makes it easier to form shocks [26].

VI. SUMMARY

We summarized several recent results on type IV bursts, type III storms, type II bursts and CME interaction, and the solar cycle variation of the number of IP type II bursts. The results show that there is a great potential for space weather applications from solar radio observations, especially at low frequencies. The results presented also demonstrate the power of multi-wavelength studies that help bring out the underlying physics, which is very important for space weather modeling.

*Acknowledgment*

I thank S. Akiyama and P. Makela for help with some figures. I also thanks the SOHO and Wind teams for making the data freely available.

*References*

[1] J. A. Klobuchar, J. M. Kunches, and A. J. VanDierendonck, "Eye on the Ionosphere: Potential Solar Radio Burst Effects on GPS Signal to Noise," GPS Solutions, vol. 3(2), pp. 69-71, 1999.

[2] A. P. Cerruti et al., "Observed solar radio burst effects on GPS/Wide Area Augmentation System carrier-to-noise ratio," Space Weather 4(10), S10006, 2006.

[3] J.-L. Bougeret, J. Fainberg, and R. G. Stone, Interplanetary radio storms. I - Extension of solar active regions through the interplanetary medium," Astron. Astrophys., vol. 136, pp. 255-262, 1984

[4] N. Gopalswamy et al., "Height of shock formation in the solar corona inferred from observations of type II radio bursts and coronal mass ejections," Adv. Space Res., vol. 51, pp. 1981-1989, 2013.

[5] J. Feynman and S. B. Gabriel, "On space weather consequences and predictions," J. Geophys. Res., vol. 105, pp. 10543-10564, 2000.

[6] N. Gopalswamy,"Interplanetary radio bursts," in Solar and Space Weather Radiophysics, Astrophysics and Space Science Library, Vol. 314, C. U. Keller and D. E. Gary, Eds., Springer Science + Business Media, Inc., 2004, pp. 305-333.

[7] N. Gopalswamy, "Coronal Mass Ejections and Solar Radio Emissions," Planetary Radio Emissions, vol. 7, pp. 325-342, 2011.

[8] G. E. Brueckner, R. A. Howard, M. J. Koomen, C. M. Korendyke, D. J. Michels, J. D. Moses, "The large angle spectroscopic coronagraph (LASCO)," Sol. Phys., vol. 162, pp. 357–402, 1995.

[9] J.-L. Bougeret et al., "Waves: The Radio and Plasma Wave Investigation on the Wind Spacecraft," Space Sci. Rev., vol. 71, pp. 231–263, 1995.

[10] Y. Leblanc, G. A. Dulk, I. H. Cairns, and J.-L. Bougeret, "Type II flare continuum in the corona and solar wind," J. Geophys. Res., vol. 105, pp. 18215–18223, 2000.

[11] N. Gopalswamy, S. Akiyama, P. Makela, S. Yashiro, and I. H. Cairns, "On the directivity of low-frequency type IV radio bursts," submitted

[12] J. Fainberg, and R. G. Stone, "Type III Solar Radio Burst Storms Observed at Low Frequencies," Sol. Phys., vol. 15, pp. 222-233, 1970.

[13] N. Gopalswamy, "Interplanetary radio bursts," in Solar and Space Weather Radiophysics, Astrophysics and Space Science Library, Vol. 314, C. U. Keller and D. E. Gary, Eds., Springer Science + Business Media, Inc., 2004, pp. 305-333.

[14] M. Reiner, M. Kaiser, M. Karlický, K. Jiřička, J. Bougeret,"Bastille Day Event: A Radio Perspective," Sol. Phys., vol. 204, pp.121-137, 2001.

[15] N. Gopalswamy, S. Yashiro, M. Kaiser, R. Howard, and J. Bougeret, "Radio Signatures of Coronal Mass Ejection Interaction: Coronal Mass Ejection Cannibalism?" Astrophys. J., vol. 548: pp. L91-L94, 2001.

[16] N. Gopalswamy, S. Yashiro, S. Krucker, G. Stenborg, and R. Howard, "Intensity variation of large solar energetic particle events associated with coronal mass ejections," J. Geophys. Res., vol. 109, p. 12105, 2004.

[17] P. K. Manoharan, N. Gopalswamy, S. Yashiro, A. Lara, G. Michalek, R. Howard, "Influence of coronal mass ejection interaction on propagation of interplanetary shocks," J. Geophys. Res., vol. 109, p. 6109, 2004.

[18] N. Gopalswamy, P. Mäkelä, S. Akiyama, S. Yashiro, H. Xie, R. J. MacDowall, M. L. Kaiser, "Radio-loud CMEs from the disk center lacking shocks at 1 AU," J. Geophys,. Res., vol. 117, p. 8106, 2012.

[19] P. Mäkelä, N. Gopalswamy, M. J. Reiner, and S. Akiyama, "Source Regions of the Type II Radio Burst Observed During a CME-CME Interaction on 2013 May 22," under preparation, 2016.

[20] L.-G. Ding et al., "Interaction between Two Coronal Mass Ejections in the 2013 May 22 Large Solar Energetic Particle Event," Astrophys. J., vol. 793, p. L35, 2014.

[21] Y. Leblanc, G. Dulk, and J.-L. Bougeret, "Tracing the Electron Density from the Corona to 1au," Sol. Phys., vol. 183, p p. 165-180, 1998.

[22] H. Cremades, F. A. Iglesias, O. C. St. Cyr, H. Xie, M. L. Kaiser, N. Gopalswamy, "Low-Frequency Type-II Radio Detections and Coronagraph Data Employed to Describe and Forecast the Propagation of 71 CMEs/Shocks," Sol. Phys., vol. 290, pp. 2455-2478, 2015.

[23] N. Gopalswamy, S. Yashiro, M. Kaiser, R. Howard, and J. Bougeret, "Characteristics of coronal mass ejections associated with long-wavelength type II radio bursts," J. Geophys. Res., vol.106, pp. 29219-29230, 2001.

[24] N. Gopalswamy et al., "Coronal mass ejections, type II radio bursts, and solar energetic particle events in the SOHO era," Annales Geophysicae, vol. 26, pp. 3033-3047, 2008.

[25] N. Gopalswamy et al.,"Large Solar Energetic Particle Events Associated with Filament Eruptions Outside of Active Regions," Astrophys. J., vol. 806, pp. 8-22, 2015.

[26] N. Gopalswamy, S. Akiyama, S. Yashiro, H. Xie, P. Makela, G. Michalek, "Anomalous expansion of coronal mass ejections during solar cycle 24 and its space weather implications," Geophys. Res. Lett., vol. 41, pp. 2673-2680, 2014.